\documentclass[a4paper,11pt]{article}
\usepackage{jcappub}

\usepackage{xcolor}
\usepackage{graphicx}
\graphicspath{%
    {./fig/}
}
\usepackage{amsmath,amssymb,amsfonts,mathrsfs,bm}
\usepackage{hyperref}
\hypersetup{%
    colorlinks=true,
    linkcolor=blue,
    citecolor=blue,
}

\newcommand{\dif}{\mathrm{d}}

\newcommand{\Mpc}{\mathrm{Mpc}}
\newcommand{\Hz}{\mathrm{Hz}}
\newcommand{\nHz}{\mathrm{nHz}}

\newcommand{\GW}{\mathrm{GW}}
\newcommand{\PBH}{\mathrm{PBH}}
\newcommand{\DM}{\mathrm{DM}}
\newcommand{\Msun}{M_\odot}
\newcommand{\rmc}{\mathrm{c}}
\newcommand{\rms}{\mathrm{s}}
\newcommand{\rmI}{\mathrm{I}}
\newcommand{\rmB}{\mathrm{B}}
\newcommand{\rmH}{\mathrm{H}}
\newcommand{\II}{\mathrm{(I)}}
\newcommand{\BB}{\mathrm{(B)}}

\begin{document}

\title{Testing primordial black hole and measuring the Hubble constant with multiband gravitational-wave observations}

\author[a,b,c]{Lang Liu,}
\author[d,e,f,*]{Xing-Yu Yang,\note[*]{Corresponding author.}}
\author[e,f,g]{Zong-Kuan Guo,}
\author[e,f,g]{Rong-Gen Cai}

\affiliation[a]{Department of Astronomy, Beijing Normal University, Beijing 100875, China}
\affiliation[b]{Advanced Institute of Natural Sciences, Beijing Normal University, Zhuhai 519087, China}
\affiliation[c]{Department of Physics, Kunsan National University, Kunsan 54150, Korea}
\affiliation[d]{Korea Institute for Advanced Study, Seoul 02455, Republic of Korea}
\affiliation[e]{CAS Key Laboratory of Theoretical Physics, Institute of Theoretical Physics, Chinese Academy of Sciences, Beijing 100190, China}
\affiliation[f]{School of Physical Sciences, University of Chinese Academy of Sciences, Beijing 100049, China}
\affiliation[g]{School of Fundamental Physics and Mathematical Sciences, Hangzhou Institute for Advanced Study, University of Chinese Academy of Sciences, Hangzhou 310024, China}

\emailAdd{liulang@bnu.edu.cn}
\emailAdd{xingyuyang@kias.re.kr}
\emailAdd{guozk@itp.ac.cn}
\emailAdd{cairg@itp.ac.cn}

\abstract{
    There exist two kinds of stochastic gravitational-wave backgrounds associated with the primordial curvature perturbations.
    One is called induced gravitational wave due to the nonlinear coupling of curvature perturbations to tensor perturbations, while the other is produced by coalescences of binary primordial black holes formed when the large amplitude curvature perturbations reenter the horizon in the radiation dominant era.
    In this paper we find a quite useful relation for the peak frequencies of these two stochastic gravitational-wave backgrounds.
    This relation can not only offer a smoking-gun criterion for the existence of primordial black holes, but also provide a method for measuring the Hubble constant $H_0$ by multiband observations of the stochastic gravitational-wave backgrounds.
}

\maketitle

\section{Introduction}

The direct detection of gravitational waves (GWs) from the merger of binary systems by LIGO/Virgo has opened a new window to explore the universe~\cite{Abbott:2016blz}.
Except for the GWs from resolvable binary coalescences, the stochastic gravitational-wave background (SGWB) produced during the evolution of the universe is also quite important.
Due to its plentiful sources, the SGWB is closely connected to many aspects of astrophysics and carries invaluable information about the evolution of the universe ranging from the late stage to the very early stage~\cite{Caprini:2018mtu, Christensen:2018iqi}.
Although the SGWB has not been detected yet, the upper limits placed on it at specific frequency bands could give important astrophysical and cosmological results~\cite{LIGOScientific:2018czr, Arzoumanian:2020vkk}.
In the future, more observations aiming at the detection of SGWB at different frequency bands will begin running.
Having the multiband observations of SGWB with the network of different experiments, one can study the inherent relation of GWs at different frequency bands and reveal information that can not be obtained by analysis on a specific band, which will lead the studies of GWs into a new stage.

Recently two bands of SGWB associated with the primordial curvature perturbations have gained increasingly broad interest.
It is well known that on CMB scales the primordial curvature perturbations are constrained tightly, however, on relatively small scales their constraints are loose~\cite{Mesinger:2005ah, Bringmann:2011ut, Chluba:2012we}.
When the primordial curvature perturbations on small scales are large enough, they can induce a sizeable GW background at second order due to the nonlinear coupling of perturbations, which could reach the sensitivity of observations~(\cite{Ananda:2006af, Baumann:2007zm, Cai:2018dig}, see review \cite{Domenech:2021ztg}, and references therein).
On the other hand, the large curvature perturbations can lead to the formation of primordial black holes (PBHs)~\cite{Zeldovich:1967lct, Hawking:1971ei, Carr:1974nx}.
PBHs have attracted considerable attention in recent years~(\cite{Saito:2008jc,Belotsky:2014kca,Carr:2016drx,Garcia-Bellido:2017mdw,Carr:2017jsz,Germani:2017bcs,Liu:2018ess,Liu:2019rnx,Fu:2019ttf,Cai:2019elf,Liu:2019lul,Cai:2019bmk,Fu:2020lob,Cai:2020fnq,DeLuca:2020sae,Vaskonen:2020lbd,DeLuca:2020agl,Domenech:2020ers,Hutsi:2020sol,Kawai:2021edk,Braglia:2021wwa,Braglia:2022icu}, see reviews~\cite{Sasaki:2018dmp,Carr:2020gox,Carr:2020xqk}, and references therein) as they are good candidates of dark matter~\cite{Carr:2016drx} and can be the sources of LIGO/Virgo detection~\cite{Bird:2016dcv,Sasaki:2016jop}, but their existence has not been confirmed yet.
With the expansion of the universe, the binary mergers of PBHs can be formed, which also produces a GW background.
Usually, these two kinds of GWs (denoted by IGWs and BGWs for short) have different characteristic frequencies and spectra, and they can be detected by multiband GW observations.

In this paper, we identify a relation of peak frequency for the spectra of IGWs and BGWs.
Such a peak frequency relation is nearly constant in a large space of parameters and can be tested by multiband observations of GWs.
The detection of GWs with such a relation will be a smoking gun for the existence of PBHs.
In addition, based on the peak frequency relation, we propose a method for measuring the Hubble constant $H_{0}$.
These results show that the joint analysis on different frequency bands of SGWB will be pretty powerful to explore some important issues in astrophysics and cosmology.
For convenience, $G=c=1$ is set throughout this paper.

\section{Multiband gravitational waves}

Considering a unimodal power spectrum of primordial curvature perturbations $\mathcal{P}_{\mathcal{R}}(k)$ which can be depicted by a set of parameters $\{ k_{*}, {\theta_1, \theta_2,\dots} \}$ (or $\{k_*, \Theta\}$ for short) as
\begin{equation}
    \mathcal{P}_{\mathcal{R}} (k) = \mathcal{P}_{\mathcal{R}} (k/k_{*}, \Theta),
\end{equation}
where $k_*$ and $\Theta$ characterize the position and the shape of the spectrum, respectively.
Such a parametrization can describe various widely used power spectra of primordial curvature perturbations, including types of delta, lognormal, broken power law and tophat, etc.

After reentering the horizon in the radiation-dominated epoch, the curvature perturbations will inevitably lead to GWs in second order due to the nonlinear coupling.
By perturbing the Einstein equations up to second order, one can get the spectrum of induced GWs\footnote{
    Usually, this spectrum is thought to depict the energy of GWs, but this is true only for GWs in linear order.
    For induced GWs, this spectrum should be understood simply as a function of the amplitude of GWs~\cite{Cai:2021jbi, Cai:2021ndu}.
}~\cite{Kohri:2018awv,Cai:2019amo},
\begin{equation}\label{eq:Omega_GW_I}
    \Omega_{\mathrm{GW}}^{\II}(k) \propto \int_{0}^{\infty} \dif v \int_{|1-v|}^{1+v} \dif u \mathcal{T} ( u,v ) \mathcal{P}_{\mathcal{R}} ( ku ) \mathcal{P}_{\mathcal{R}} ( kv ),
\end{equation}
where $\mathcal{T}(u,v)$ is a function of $u$ and $v$.
The spectrum of GWs induced by curvature perturbations with unimodal power spectrum has a peak whose position is given by
\begin{equation}\label{eq:pk_I}
    \nu_{\rmI} \equiv k_{\rmI}/(2\pi) = k_{*}~C_{\rmI}(\Theta ),
\end{equation}
where $\nu_{\rmI}$ denotes the peak frequency of IGWs and $C_{\rmI}(\Theta)$ is a function of the parameter set $\Theta$.

On the other hand, when the primordial curvature perturbations are large enough, they can lead to the formation of PBHs.
PBHs could form a binary in the early universe and coalesce within the age of the universe.
The spectrum of GWs produced from coalescing PBH binaries is given by~\cite{Wang:2016ana, Wang:2019kaf}
\begin{equation}\label{eq:Omega_GW_B}
    \Omega_{\GW}^{\BB}(\nu) \propto \nu \int \dif M_{1} \dif M_{2} \int_{0}^{\frac{\nu_{\rmc}}{\nu}-1} \dif z \frac{R(z,M_{1},M_{2})}{(1+z)H(z)} \frac{\dif E_{\mathrm{GW}}}{\dif \nu_{\rms}},
\end{equation}
where $M_{1}$ and $M_{2}$ are the masses of PBHs, $R(z,M_{1},M_{2})$ is the merger rate distribution of PBH binaries~\cite{Raidal:2018bbj}, $H(z)$ is the Hubble parameter, $\frac{\dif E_{\GW}}{\dif\nu_{\rms}}$ is GW energy spectrum of a binary black hole (BBH) coalescence, $\nu_\rms$ is the frequency in the source frame which is related to the observed frequency $\nu$ through $\nu_\rms=(1+z)\nu$, and $\nu_{\rmc}$ is the cutoff frequency for a given BBH system.
The spectrum of GWs from binary PBH mergers depends on the power spectrum of the curvature perturbations by the relation
$\Omega_{\GW}^{\BB} \Leftarrow R(z,M_{1},M_{2}) \Leftarrow f(M) \Leftarrow \mathcal{P}_{\mathcal{R}}$, where $f(M)\equiv\frac{1}{\Omega_{\DM}}\frac{\dif \Omega_{\PBH}}{\dif \ln M}$ is the mass function of PBHs.

By making use of the Press-Schechter formalism, the mass function of PBH can be obtained as
\begin{equation}
    f(M) = 2\frac{\Omega_{\mathrm{m}}}{\Omega_{\DM}} \int_{-\infty}^{\infty} \frac{a_{\mathrm{eq}}}{a_{M_{\rmH}}} \tilde{\beta}_{M_{\rmH}} (M) \dif \ln M_{\rmH},
\end{equation}
where
\begin{equation}
    \begin{aligned}
        \tilde{\beta}_{M_{\rmH}}(M) = \frac{K}{\gamma \sqrt{2\pi\sigma^{2}}} \left( \frac{M}{K M_{\rmH}} \right)^{1+\frac{1}{\gamma}}
        \exp \left( -\frac{1}{2 \sigma^2 } \left[ \left( \frac{M}{K M_{\rmH}} \right)^{\frac{1}{\gamma}} + \delta_{\rmc} \right]^2\right),
    \end{aligned}
\end{equation}
and
\begin{equation}
    \sigma^{2}(k) = \int_{0}^{\infty} (dq/q) \tilde{W}^{2}(k^{-1}, q)\frac{16}{81}(q/k)^{4} T^{2}(k^{-1}, q) \mathcal{P}_{\mathcal{R}}(q)
\end{equation}
is the variance of density contrast, where $\tilde{W}=\exp(-q^{2}/k^{2}/2)$ is the Fourier transform of a volume-normalized Gaussian window smoothing function, and $T=3(\sin l - l \cos l)/l^{3}$ with $l \equiv q/k/\sqrt{3}$ is the transfer function.
The critical collapse of an overdensity region gives the relation of the PBH mass $M$, and the density contrast $\delta$ as $M= M_{\rmH}K(\delta-\delta_{\rmc})^{\gamma}$, where $M_{\rmH}$ is the horizon mass, $K=3.3$, $\gamma=0.36$ and $\delta_{\rmc}=0.45$ are numerical constants\footnote{
    Dependence of $\delta_{\rmc}$ on the radial profile of the density perturbations is neglected~\cite{Musco:2018rwt}.
}.

It is obvious that the dependence of $\sigma^{2}$ on the parameter set $\{k_{*}, \Theta \}$ can be depicted by $\sigma^{2}(k)=\sigma^{2}(x, \Theta)$, where the dimensionless quantity $x$ is defined as $x \equiv k/k_{*}$.
Considering the evolution of entropy density and the Friedmann equation in the radiation-dominated era, one has
\begin{equation}
    a_{M_{\rmH}} \propto M_{\rmH}^{1/2} \propto k^{-1},
\end{equation}
which gives $a_{M_{\rmH}}/a_{M_{*}} = (M_{\rmH}/M_{*})^{1/2} = (k/k_{*})^{-1} = x^{-1}$, where $M_{*} \equiv M_{\rmH}(k_{*})$.
Denoting $s \equiv M/M_{*}$, one can find the dependence of $\tilde{\beta}_{M_{\rmH}}$ on the parameter set $\{k_{*}, \Theta \}$ can be depicted by $\tilde{\beta}_{M_{\rmH}}(M) = \tilde{\beta}(s,x,\Theta)$.
Thus one can obtain
\begin{equation}
    f(M) = 2\frac{\Omega_{\mathrm{m}}}{\Omega_{\DM}} \frac{a_{\mathrm{eq}}}{a_{M_{*}}} \int_{0}^{\infty} 2 \tilde{\beta}(x,s,\Theta) \dif x .
\end{equation}

In the nonspinning limit, the inspiral-merger-ringdown energy spectrum for a BBH coalescence can be rewritten as~\cite{Ajith:2007kx,Ajith:2009bn}
\begin{equation}
    \frac{\dif E_{\mathrm{GW}}}{\dif \nu_{\rms}} \propto \tilde{E}(y,z,s_{1},s_{2}),
\end{equation}
where $y \equiv \nu M_{*}$.
Notice that $\dif M_{1} \dif M_{2} R(z,M_{1},M_{2}) \propto \dif s_{1} \dif s_{2} \tilde{R}(z,s_{1},s_{2},\Theta)$, then Eq.~\eqref{eq:Omega_GW_B} can be rewritten as
\begin{equation}
    \begin{aligned}
        \Omega_{\GW}^{\BB}(\nu) \propto
        y \int \dif s_{1} \dif s_{2} \int_{0}^{y_{\rmc}(s_{1},s_{2})/y-1} \dif z \frac{\tilde{R}(z,s_{1},s_{2},\Theta)}{(1+z)H(z)} \tilde{E}(y,z,s_{1},s_{2}).
    \end{aligned}
\end{equation}
One can find that there is a peak in this spectrum of GWs when the spectrum of the curvature perturbations is unimodal, and the peak frequency is given by
\begin{equation} \label{eq:pk_B}
    \nu_{\rmB} \equiv M_{*}^{-1} y_{\rmB} = M_{*}^{-1}~C_{\rmB}(\Theta) ,
\end{equation}
where $C_{\rmB}$ is a function of parameter $\Theta$.

When detecting GWs, normally the frequency with a larger amplitude is easier to be detected.
Therefore the peak frequencies $\nu_{\rmI}$ and $\nu_{\rmB}$ are most likely to be detected in multiband observations, and they contain information on the primordial curvature perturbations and PBHs due to their dependence on the parameter set $\{ k_{*},\Theta \}$.

\section{Peak frequency relation}

By definition, the horizon mass is $M_{\rmH}=\frac{4 \pi}{3} \rho H^{-3}$.
Using the scaling of the energy density with the temperature during the radiation-dominated era $\rho \propto g_{* r}(T) T^{4}$, and the conservation of entropy $g_{* s}(T) T^{3} a^{3}=\text{constant}$, one can obtain
\begin{equation}
    M_{\rmH}=k^{-2} H_{0} \Omega_{r,0}^{1/2} \frac{1}{2} (\frac{g_{*s}(T)}{g_{*s}(T_0)})^{-\frac{2}{3}}(\frac{g_{*r}(T)}{g_{*r}(T_0)})^{\frac{1}{2}},
\end{equation}
where $\Omega_{r,0}$ denotes the fraction of radiation energy at the present time, $g_{*s}$ and $g_{*r}$ are the effective degrees of freedom for the entropy density and the radiation respectively.
Combining Eqs.\eqref{eq:pk_I} and \eqref{eq:pk_B}, we arrive at the relation of the two peak frequencies
\begin{equation} \label{eq:peak_relation}
    \begin{aligned}
        \frac{\nu_{\rmI}^{2}}{\nu_{\rmB}} &= H_{0} \Omega_{r,0}^{1/2} \left[ \frac{1}{2} (\frac{g_{*s}(T_*)}{g_{*s}(T_0)})^{-\frac{2}{3}}(\frac{g_{*r}(T_*)}{g_{*r}(T_0)})^{\frac{1}{2}} \right] \left[ \frac{C_{\rmI}(\Theta)^{2}}{C_{\rmB}(\Theta)} \right] \\
                                          &\equiv H_{0}\Omega_{r,0}^{1/2}~G(k_{*})~C(\Theta) \equiv H_{0} \Omega_{r,0}^{1/2}~Y(k_{*},\Theta),
    \end{aligned}
\end{equation}
where $T_*$ is temperature when the wavenumber $k_*$ reenters the Hubble horizon.

\begin{figure}[htbp]
    \centering
    \includegraphics[width=0.7\textwidth]{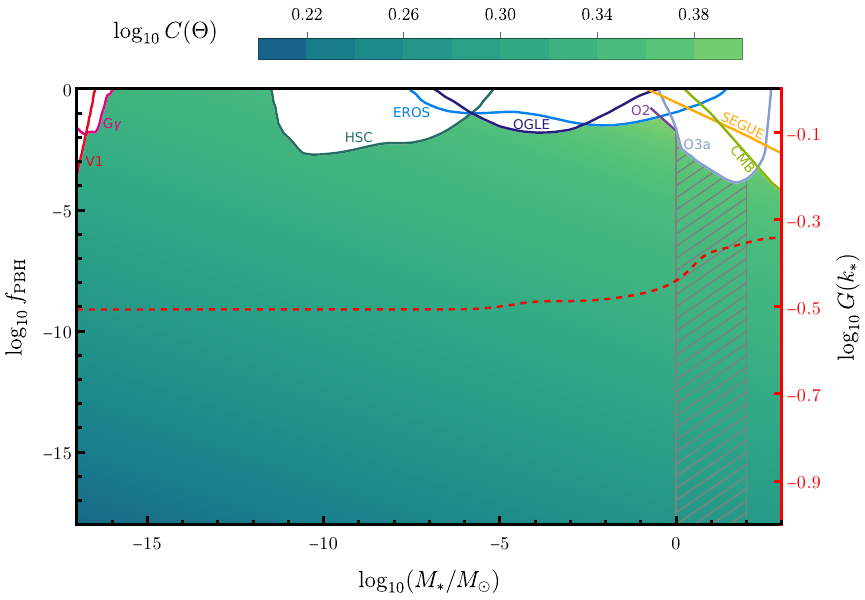}
    \caption[]{Plot of $G(k_{*})$ and $C(\Theta)$ in the $M_{*}-f_{\PBH}$ plane, in which colorized region denotes $\log_{10}C(\Theta)$ and the red dashed line denotes $\log_{10}G(k_{*})$.
        Some regions are excluded due to the constraints on PBH abundance from the galactic 511 keV line from Hawking radiation (G$\gamma$ \cite{DeRocco:2019fjq, Laha:2019ssq, Dasgupta:2019cae, Laha:2020ivk}), Voyager 1 measurements (V1 \cite{Boudaud:2018hqb}), gravitational lensing events (HSC \cite{Niikura:2017zjd}, EROS \cite{EROS-2:2006ryy}, OGLE \cite{Niikura:2019kqi}), LIGO/Virgo observations (O2 \cite{LIGOScientific:2019kan}, O3a \cite{Hutsi:2020sol}), dynamical effects (SEGUE \cite{Koushiappas:2017chw}), and cosmic microwave background (CMB~\cite{Poulin:2017bwe}).
    }
    \label{fig:lgC_lgG_delta}
\end{figure}

We focus on PBHs with mass in the range $[10^{-17},10^{3}] \Msun$, which contains the mass window for LIGO/Virgo detection and the mass window in which PBHs can constitute all dark matter.
For horizon mass $M_{*} \sim [10^{-17},10^{3}] \Msun$, the corresponding position parameter is $k_{*}\sim [10^{15},10^{5}] \Mpc^{-1} \sim [10^{10},1] \nHz$.
The position parameter $k_*$ influences the peak frequency relation through the factor $G(k_{*})$ which depends on the thermal history of the universe, and one can find that $G(k_{*})$ is nearly a constant~\cite{Saikawa:2018rcs}.
As to the influence of the shape parameters $\Theta$ on the peak frequency relation, we investigate the delta spectrum $\mathcal{P}_{\mathcal{R}}(k)=A\ \delta( \ln(k/k^{*}) )$ firstly.
In this case, the shape parameter is $\Theta=\{A\}$.
Since the parameter $A$ affects the PBH abundance $f_{\PBH}$ exponentially, but affects the $C(\Theta)$ polynomially, one can anticipate that $C(\Theta)$ is nearly a constant even $f_{\PBH}$ varies in a huge range.
In figure~\ref{fig:lgC_lgG_delta} we show the numerical value of $\log_{10}G(k_{*})$ and $\log_{10}C(\Theta)$, which verifies the anticipation.
Combining the result of $G(k_{*})$ and $C(\Theta)$, one has $Y(k_{*},\Theta)\sim 1$ for PBHs with mass in the range $[10^{-17},10^{3}] \Msun$ and abundance in the range $[10^{-18},1]$.
This indicates that even though the mass of PBHs varies in a considerable range (20 orders of magnitude), the peak frequency relation is nearly constant.
In other words, one always has
\begin{equation}\label{eq:peak_relation_nearly}
    \frac{\nu_{\rmI}^{2}}{\nu_{\rmB}} \sim H_{0} \Omega_{r,0}^{1/2}
\end{equation}
for $\nu_{\rmI} \sim [10^{-10},1]\Hz$ and $\nu_{\rmB} \sim [1,10^{20}]\Hz$.

\begin{figure}[htbp]
    \centering
    \includegraphics[width=0.7\textwidth]{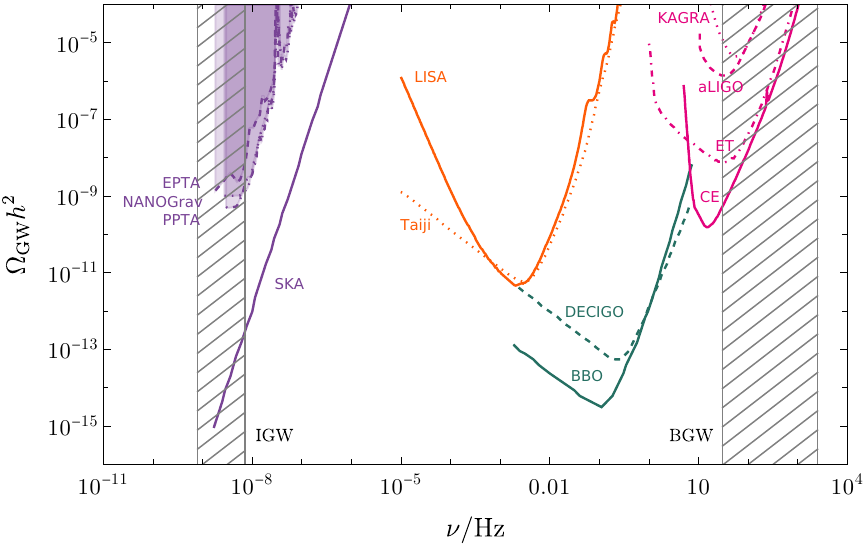}
    \caption{Peak frequency regions for GWs corresponding to PBHs with mass in $[10^{0},10^{2}] \Msun$.
    The constrain or the sensitivity curves of GW detectors including: EPTA~\cite{Lentati:2015qwp}, NANOGrav~\cite{Arzoumanian:2018saf}, PPTA~\cite{Shannon:2015ect}, SKA~\cite{Carilli:2004nx}, LISA~\cite{LISA:2017pwj}, Taiji~\cite{Guo:2018npi}, DECIGO~\cite{Kawamura:2011zz}, BBO~\cite{Phinney.NMCS.2004.}, KAGRA~\cite{Somiya:2011np}, aLIGO~\cite{LIGOScientific:2014pky}, ET~\cite{Punturo:2010zz}, CE~\cite{LIGOScientific:2016wof}. }
    \label{fig:IGWBGWParaSpace}
\end{figure}

Although the peak frequency relation is valid in a large parameter space, it is hard to detect GWs with peak frequency in the whole parameter space due to the limitation of current experiments.
In figure~\ref{fig:IGWBGWParaSpace}, the slash regions show the range of peak frequency for IGW and BGW corresponding to PBH with mass in $[10^{0},10^{2}] \Msun$.
Considering the development of experimental observations, GWs in such regions are most likely to be detected in the near future.

\begin{figure}[htpb]
    \centering
    \includegraphics[width=0.7\textwidth]{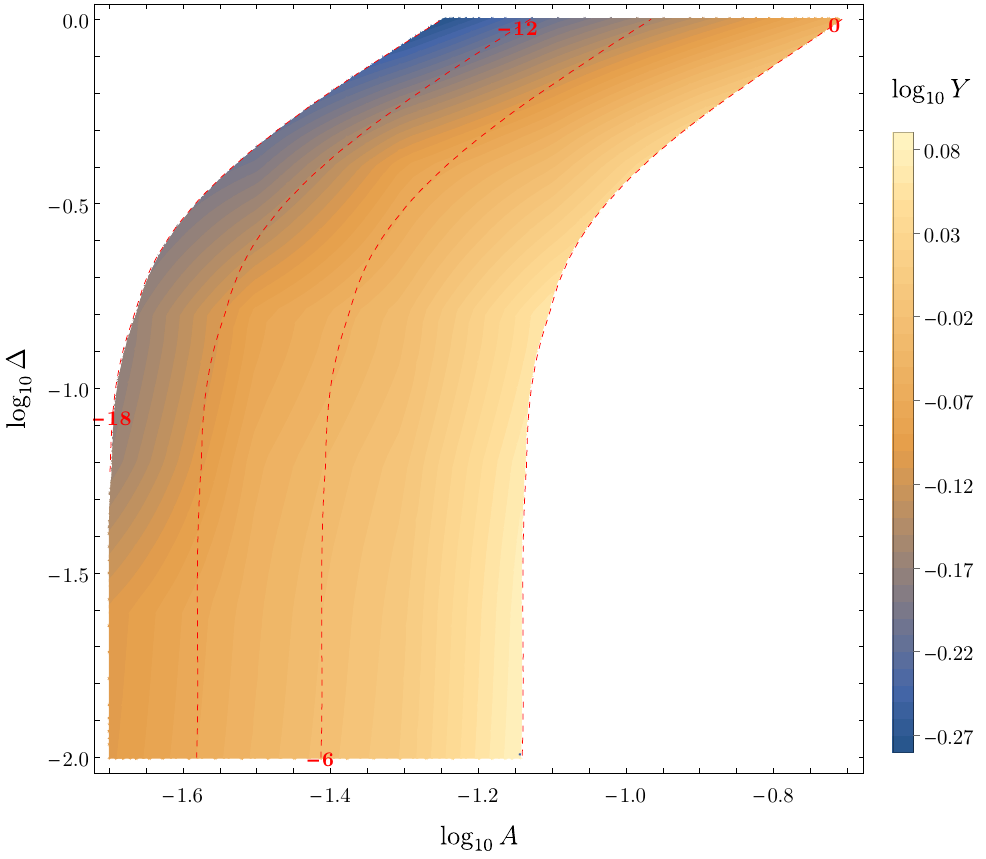}
    \caption[]{The numerical value of $Y$ in the $\Delta-A$ plane with $M_*=10 \Msun$. The red dashed lines correspond to the contour lines of $\log_{10}f_{\PBH}$.}
    \label{fig:cp$lgY}
\end{figure}

Usually, the lognormal power spectrum of primordial curvature perturbations
$\mathcal{P}_{\mathcal{R}} (k) = \frac{A}{\sqrt{2\pi}\Delta} \exp \left( -\frac{\ln^{2}(k/k_{*})}{2\Delta^{2}} \right)$
is a good fit for a peaked power spectrum around its peak, and it appears naturally in many different inflation models~\cite{Pi:2020otn}.
As shown in figure~\ref{fig:cp$lgY}, one can find that $Y(k_{*},\Theta) \sim \mathcal{O}(1)$ no matter there is a narrow peak $(\Delta \ll 1)$ or a broad peak $(\Delta \sim 1)$ in the spectrum of curvature perturbations.
In the mass range $[10^{-17},10^{3}] \Msun$, which contains the mass window for LIGO/Virgo detection and the mass window in which PBHs can constitute all dark matter, we find $Y(k_{*},\Theta)$ is always around $\mathcal{O}(1)$ for the widely used power spectra of primordial curvature perturbations, including types of delta, lognormal, broken power law and tophat.
Therefore, in future multiband GW observations, if two different GWs with peak frequencies satisfying $\nu_{\rmI}^{2}/\nu_{\rmB} \sim H_{0} \Omega_{r,0}^{1/2}$ are detected, one can confirm the existence of PBHs.
In addition, our calculations hint that for all the unimodal power spectrum of primordial curvature perturbations with moderate peak width, the peak frequency relation Eq.\eqref{eq:peak_relation_nearly} is valid.

Recently, the 12.5-year pulsar timing data set released by the North American Nanohertz Observatory for Gravitational Waves (NANOGrav) shows evidence of a stochastic common-spectrum process~\cite{Arzoumanian:2020vkk}.
The importance of the quadrupole nature in the overlap reduction function is inconclusive, while monopole and dipole are relatively disfavored, which means that the NANOGrav collaboration may have detected a stochastic GW background with astrophysical or cosmological origin~\cite{Bian:2021lmz}.
If the signal of NANOGrav is caused by induced GWs, i.e., $\nu_{\rmI} \sim 10^{-8.3}\Hz$, there is inevitably a corresponding GW signal with $\nu_{\rmB} \sim 10^{3}\Hz$ produced by merger events of binary solar-mass PBHs, which can be detected by future GW observations~\cite{Kohri:2020qqd}, and therefore offering a criterion for the existence of PBHs.

\begin{figure}[htbp]
    \centering
    \includegraphics[width=0.48\textwidth]{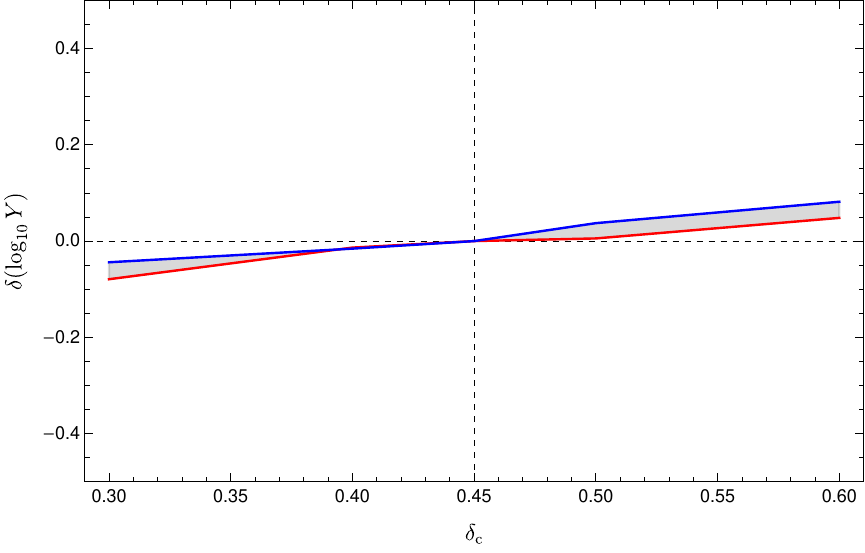}
    \includegraphics[width=0.48\textwidth]{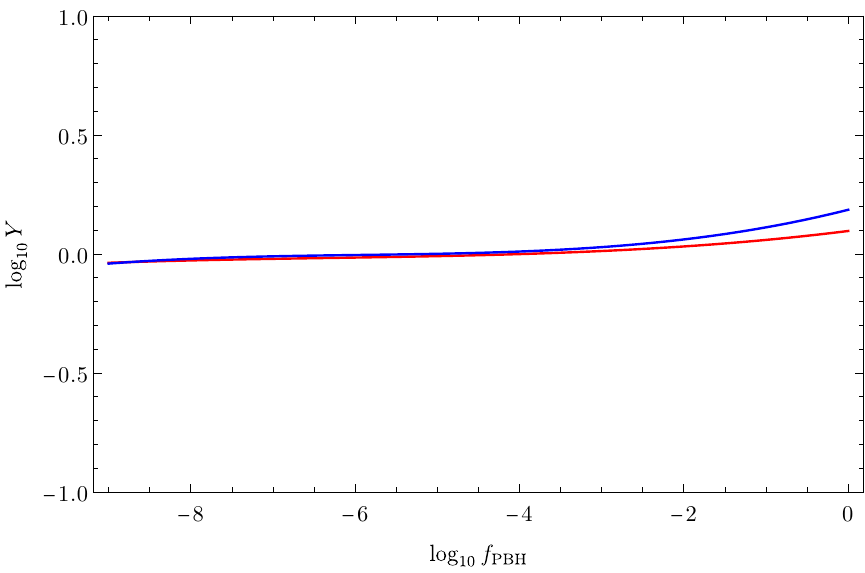}
    \caption[]{\textit{Left:} Differences of $Y$ with different threshold $\delta_{\rmc}$ compared to $Y$ with $\delta_{\rmc}=0.45$ in the ranges $M_{\PBH} \in [10^{0},10^{2}] \Msun$ and $f_{\PBH} \in [10^{-8}, 10^{-4}]$ where the red (blue) line denotes the left (right) boundary.
        \textit{Right:} Factor $Y$ as function of $f_{\PBH}$ for different suppression factor $S$ when $M_{*}=10 \Msun$.
        The red line represents the case with a simple form $S$ or no $S$~\cite{Liu:2018ess}.
        The blue line represents the case with a complicated form $S$~\cite{Hutsi:2020sol}.
    }
    \label{fig:lgY}
\end{figure}

This criterion for the existence of PBHs has several advantages:
it weakly depends on PBH abundance which has a fine-tuning problem;
it is weakly affected by the uncertainties of PBH mass function caused by the different choices of the window function, threshold $\delta_{\rmc}$, the transfer function of curvature perturbations, since these uncertainties influence the peak frequency relation only through the factor $C(\Theta)$ which remains almost unchanged.
In the left panel of figure~\ref{fig:lgY}, we show the differences of $Y$ with different threshold $\delta_{\rmc}$ compared to $Y$ with $\delta_{\rmc}=0.45$ in the ranges $M_{\PBH} \in [10^{0},10^{2}] \Msun$ and $f_{\PBH} \in [10^{-8}, 10^{-4}]$ where the red (blue) line denotes the left (right) boundary, which explicitly shows that $Y$ is weakly affected by the threshold $\delta_{\rmc}$.
The suppression factor in the merger rate is chosen to be $S=\left(1+\sigma_{\mathrm{M}}^2 / f_{\mathrm{PBH}}^2\right)^{-21 / 74}$ in this work.
In such a case, the suppression factor only changes the merger rate and does not change the value of factor $Y$.
A accurate suppression factor $S$ with a complicated form is derived in~\cite{Hutsi:2020sol} (Eqs.~(2.2) and (2.7)).
The right panel of figure~\ref{fig:lgY} shows $Y$ as a function of $f_{\PBH}$ for different suppression factors $S$.
The red line represents the case with a simple form $S$ or no $S$~\cite{Liu:2018ess}, and the blue line represents the case with a complicated form $S$~\cite{Hutsi:2020sol}.
Therefore, one can find the suppression factor $S$ also weakly affects the constancy of function Y.

\section{Hubble constant}

The Hubble constant $H_{0}$ measured from Planck~\cite{Planck:2018vyg} is inconsistent with that measured from the Supernovae H0 for the Equation of State (SH0ES) project~\cite{Riess:2021jrx} up to $5\sigma$ level.
Whether this Hubble tension is caused by measurement or it indicates physics beyond the $\Lambda$CDM model is still under debate.
Therefore a measurement of $H_{0}$ independent of the cosmic distance ladder may shed some light on this problem.
GWs from compact binary coalescences allow for the direct measurement of the luminosity distance to their source, which makes them standard-distance indicators.
In conjunction with an electromagnetic counterpart or an identified host galaxy, GWs can be used as standard sirens to construct a distance-redshift relation and measure the Hubble constant~\cite{LIGOScientific:2017adf, LIGOScientific:2019zcs}.
However, binary coalescences with detectable electromagnetic counterparts are rare, and identifying the host galaxy is also not accessible due to the current sensitivity of detectors.

Based on the peak frequency relation Eq.~\eqref{eq:peak_relation}, one can find a new method to measure the Hubble constant, which is given by
\begin{equation}\label{H0}
    H_{0} = \Omega_{r,0}^{-1/2} \frac{\nu_{\rmI}^{2}}{\nu_{\rmB}}\ Y(k_{*},\Theta)^{-1}.
\end{equation}
Such a method does not depend on the distance-redshift relation, therefore only GW observations are needed.
Through multiband GW observations, if two stochastic GW backgrounds satisfying $\nu_{\rmI}^{2}/\nu_{\rmB} \sim H_{0} \Omega_{r,0}^{1/2}$ are detected, one can obtain the parameter set $\{k_{*},\Theta\}$ by data fitting.
After that, one can calculate the factor $Y(k_{*},\Theta)$ and obtain the Hubble constant.

The uncertainties in $H_{0}$ can be divided into two different parts.
One is from the experiments including measurement uncertainties of the peak frequency and data fitting uncertainties of factor $Y$.
The other is the model uncertainty which depends on the details of PBH formation.
The different detail of critical collapse doesn't change the form of Eq.~\eqref{H0}, but it changes the value of $Y$ for a given parameter set $\{k_{*},\Theta\}$.

For the experimental uncertainty, according to the propagation of uncertainty, one has
\begin{equation}
    \left(\frac{\sigma_{H_{0}}}{H_{0}}\right)^{2} \approx\left(\frac{\sigma_{\nu_{\rmI}}}{\nu_{\rmI}}\right)^{2}+\left(\frac{\sigma_{\nu_{\rmB}}}{\nu_{\rmB}}\right)^{2}+\left(\frac{\sigma_{Y}}{Y}\right)^{2}+\left(\frac{\sigma_{\Omega_{r,0}}}{\Omega_{r,0}}\right)^{2},
\end{equation}
where $\sigma_{\dots}$ represents the corresponding standard deviation.
In the future, if two different GWs which satisfy the peak frequency relation could be detected, the peak frequencies of two different stochastic GW backgrounds can be measured extremely precisely.
Therefore, the main uncertainty is from the data fitting of factor $Y$.
And with the development of experimental detections, the more information about the two different SGWBs we know, the less uncertainty of factor $Y$ we get.

\begin{figure}[htbp]
    \centering
    \includegraphics[width=0.7\textwidth]{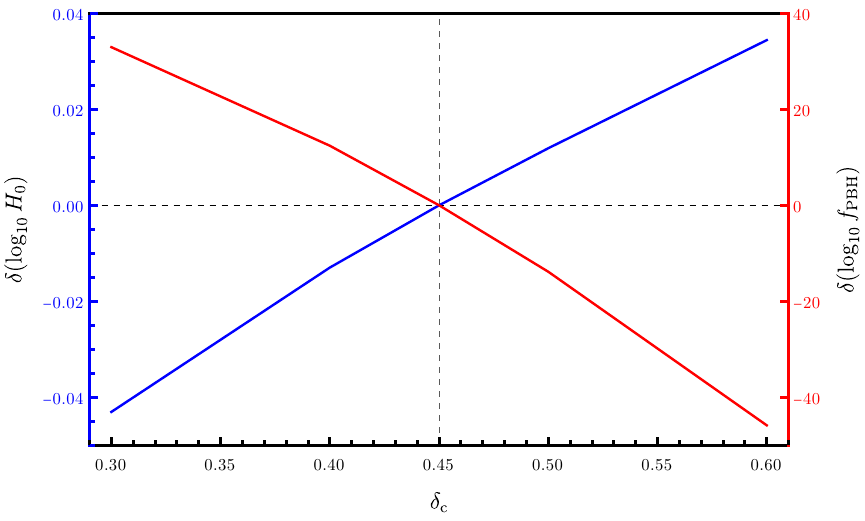}
    \caption[]{Differences of the estimated value of $\log_{10}H_{0}$ and $\log_{10}f_{\PBH}$ compared with $\delta_{\rmc}=0.45$.
        Here we set $A=0.008$ and $M_{*}=10\Msun$ for demonstration.
    }
    \label{fig:deltalgH0_deltalgfPBH}
\end{figure}

For the model uncertainty, one needs the simulation of critical collapse to confirm the detail of PBH formation.
However, in figure~\ref{fig:deltalgH0_deltalgfPBH}, we show that the detail of PBH formation such as the threshold $\delta_{\rmc}$ for PBH formation can affect the PBH abundance significantly, but it barely affects the estimated value of $H_{0}$.
Further studies on this new method will be done in the future.

\section{Conclusion}

When the primordial curvature perturbations on small scales are large enough, they can lead to two kinds of GWs which can be detected by multiband GW observations.
One is GWs induced by the primordial curvature perturbations at second order in the radiation-dominated era, the other is GWs emitted by coalescences of binary PBHs associated with the primordial curvature perturbations.
In this paper, we demonstrated for the first time that the relation of peak frequencies for the spectra of these two SGWBs is $\nu_{\rmI}^{2}/\nu_{\rmB} \sim H_{0} \Omega_{r,0}^{1/2}$.
This peak frequency relation offers a new criterion for the existence of PBHs, and the detections of GWs with such a relation will be a smoking gun of the existence of PBHs.

We note that there are many studies revealing that the spectrum of SGWB presents a peak feature with an enhanced amplitude that can be at reach for current and forthcoming terrestrial and space-based GWs surveys.
And this peak relation is valid for all these frequency ranges.
Especially, assuming the signal detected by NANOGrav is caused by the induced SGWB, we show that if GWs whose spectrum peaks around $10^{3}\Hz$ are detected in the future, one could confirm the existence of PBHs.

Our work shed new light on the measurement of the Hubble constant.
The peak frequency relation provides a new method for measuring $H_0$ through multiband observations of the SGWB.
Our analysis differs substantially from those on Planck or other measurements in the literature, which were based on the standard siren method.
The key novel feature in our method is that the Hubble constant appears in the Horizon mass.
Therefore, such a method only needs GW data and does not need the redshift information which is necessary for the standard siren.

\section*{Acknowledgments}

This work is supported in part by the National Key Research and Development Program of China Grant No.2020YFC2201502, and by the National Natural Science Foundation of China Grants No.11690022, No.11821505, No.11991052, No.11947302, No.12235019, and by the Strategic Priority Research Program of the CAS Grant No.XDPB15, and by the Key Research Program of Frontier Sciences of CAS.
The work of LL is supported in part by the National Research Foundation of Korea (NRF) funded by the Ministry of Education (2019R1I1A3A01063183) and by the National Natural Science Foundation of China No.12247112 and No.12247176.
The work of XYY is supported in part by the KIAS Individual Grant QP090701.

\appendix

\section{Relation between the horizon mass $M_{\rmH}$ and the wavenumber $k$ during the radiation-dominated era}

According to the conservation of entropy,
\begin{equation}
    g_{* s}(T) T^{3} a^{3}=\text {constant},
\end{equation}
the scale factor $a$ can be given by
\begin{equation} \label{Aa}
    a = \left( \frac{g_{*s}(T_0)}{g_{*s}(T)} \right)^{1/3} \frac{T_0}{T},
\end{equation}
where $a_{0}=1$ is used, $g_{*s}$ is the effective degrees of freedom for the entropy density and $T_{0}=2.725\mathrm{K}$ is the present temperature of the cosmic microwave background.
From the Friedmann equation and the scaling of the energy density with the temperature during the radiation-dominated era,
\begin{equation}
    \frac{3 H^2}{8\pi}= \rho \approx \rho_{r} = \frac{\pi^2 g_{*r} (T)}{30} T^4\ ,
\end{equation}
where $g_{*r}$ the effective degrees of freedom for the radiation, one has
\begin{equation} \label{AH}
    H=\left(\frac{8\pi}{3} \frac{\pi^2 g_{*r} (T)}{30} \right)^{1/2} T^2.
\end{equation}
Combining Eqs.~\eqref{Aa} and \eqref{AH}, by definition, one has
\begin{equation}
    k=aH=\left(\frac{g_{*s}\left(T_0\right)}{g_{*s}(T)}\right)^{1/3} T_0\left(\frac{8 \pi}{3} \frac{\pi^2}{30} g_{*r}(T)\right)^{1/2} T.
\end{equation}
From the definition of the horizon mass, $M_{\rmH}=\frac{4 \pi}{3} \rho H^{-3}$, the relation between the horizon mass $M_{\rmH}$ and the wavenumber $k$ is given by
\begin{equation}
    \begin{aligned}
        M_{\rmH}
        &=\frac{4 \pi}{3} \frac{3}{8 \pi} H^2 H^{-3}=\frac{1}{2} H^{-1}=\frac{1}{2}\left(\frac{8 \pi}{3} \frac{\pi^2}{30} g_{* r}(T)\right)^{-1/2} T^{-2} \\
        &=\frac{1}{2}\left(\frac{8 \pi}{3} \frac{\pi^2}{30} g_{* r}(T)\right)^{-1/2} k^{-2}\left[ \left(\frac{g_{* s}\left(T_0\right)}{g_{*s}(T)}\right)^{1/3} T_0\left(\frac{8 \pi}{3} \frac{\pi^2}{30} g_{* r}(T)\right)^{1/2}\right ]^2\\
        &=k^{-2} \frac{1}{2}\left(\frac{g_{*s}\left(T_0\right)}{g_{*s}(T)}\right)^{2/3}\left(\frac{g_{* r}(T)}{g_{*r}\left(T_0\right)}\right)^{1/2}\left(\frac{8 \pi}{3} \frac{\pi^2}{30} g_{* r}\left(T_0\right) T_0^4\right)^{1/2}\\
        &=k^{-2} \frac{1}{2}\left(\frac{g_{* s}\left(T_0\right)}{g_{*s}(T)}\right)^{2/3}\left(\frac{g_{*r}(T)}{g_{* r}\left(T_0\right)}\right)^{1/2}\left(\frac{8 \pi}{3} \rho_{r, 0}\right)^{1/2}\\
        &=k^{-2} \frac{1}{2}\left(\frac{g_{* s}\left(T_0\right)}{g_{*s}(T)}\right)^{2/3}\left(\frac{g_{*r}(T)}{g_{* r}\left(T_0\right)}\right)^{1/2} H_0 \Omega_{r,0}^{1/2},
    \end{aligned}
\end{equation}
where $H_0$ is Hubble constant and $\Omega_{r,0} \equiv \rho_{r,0}/\rho_{c,0} $ denotes the fraction of radiation energy at present time.

\section{Spectrum of GWs produced from coalescing PBH binaries}

The spectrum of GWs produced from coalescing PBH binaries is given by~\cite{Wang:2016ana, Wang:2019kaf}
\begin{equation}
    \Omega_{\GW}^{\BB}(\nu) = \frac{\nu}{\rho_c} \int \dif M_{1} \dif M_{2} \int_{0}^{\frac{\nu_{\rmc}}{\nu}-1} \dif z \frac{R(z,M_{1},M_{2})}{(1+z)H(z)} \frac{\dif E_{\mathrm{GW}}}{\dif \nu_{\rms}},
\end{equation}
where $\rho_c$ is the critical density, $M_{1}$ and $M_{2}$ are the masses of PBHs, $R(z,M_{1},M_{2})$ is the merger rate distribution of PBH binaries~\cite{Raidal:2018bbj}, $H(z)$ is the Hubble parameter, $\frac{\dif E_{\GW}}{\dif\nu_{\rms}}$ is GW energy spectrum of a BBH coalescence, $\nu_\rms$ is the frequency in the source frame which is related to the observed frequency $\nu$ through $\nu_\rms=(1+z)\nu$, and $\nu_{\rmc}$ is the cutoff frequency for a given BBH system.

In the non-spinning limit, the inspiral-merger-ringdown energy spectrum for a BBH coalescence is \cite{Ajith:2007kx,Ajith:2009bn}
\begin{equation}\label{dEdnus}
    \frac{dE_{\mathrm{GW}}}{d\nu_{\mathrm{s}}}(\nu_{\mathrm{s}})=\frac{\pi^{2/3}M_{c}^{5/3}}{3}
    \begin{cases}
        \nu_{\mathrm{s}}^{ -1/3}, &\text{for}~\nu_{\mathrm{s}}<\nu_1 \\
        w_1 \nu_{\mathrm{s}}^{2/3}, &\text{for}~\nu_1\le \nu_{\mathrm{s}}<\nu_2\\
        w_2\frac{\sigma^{4}\nu_{\mathrm{s}}^{2}}{\left(\sigma^{2}+4\left(\nu_{\mathrm{s}}-\nu_{2}\right)^{2}\right)^{2}}, &\text{for}~\nu_2\le \nu_{\mathrm{s}}\le \nu_3\\
        0, &\text{for}~\nu_3 < \nu_{\mathrm{s}}
    \end{cases}
\end{equation}
where $M_{c}^{5/3}=M_{1}M_{2}/(M_{1}+M_{2})^{1/3}$ is the chirp mass, $w_1=\nu_1^{-1}$ and $w_2=\nu_1^{-1}\nu_2^{-4/3}$ are two constants which make the energy spectrum to be continuous.
The parameters $\sigma$ and $\nu_i$ ($i=1,2,3$) could be expressed in terms of the total mass $M_{t}=M_{1}+M_{2}$ and the symmetric mass ratio $\eta=M_{1}M_{2}/(M_{1}+M_{2})^{2}$ as
\begin{align}
&\pi M_{t} \nu_1=(1-4.455+3.521)+0.6437\eta-0.05822\eta^2-7.092\eta^3,\\
&\pi M_{t} \nu_2 = (1-0.63)/2+0.1469\eta-0.0249\eta^2+2.325\eta^3,\\
&\pi M_{t} \sigma = (1-0.63)/4 -0.4098\eta +1.829\eta^2-2.87\eta^3,\\
&\pi M_{t} \nu_3 = 0.3236 -0.1331\eta -0.2714\eta^2 +4.922\eta^3.
\end{align}
Recalling the definition that $s \equiv M/M_{*}$, one has $M_t=(s_1+s_2)M_*$ and $\eta=s_1s_2/({s_1+s_2})^{2}$, then \eqref{dEdnus} can be rewritten as
\begin{equation} \label{dEdnusM}
    \begin{aligned}
        \tilde{E}(\nu_{\mathrm{s}}M_*,s_{1},s_{2}) &\equiv \frac{dE_{\mathrm{GW}}}{d\nu_{\mathrm{s}}}(\nu_{\mathrm{s}}) =\frac{\pi^{2/3}M_{*}^{2}s_1 s_2(s_1+s_2)^{-1/3}}{3} \\
                                                   &\times
                                                   \begin{cases}
                                                       (\nu_{\mathrm{s}}M_*)^{ -1/3}, &\text{for}~\nu_{\mathrm{s}}M_*<\nu_1 M_*\\
                                                       (\nu_1 M_*) (\nu_{\mathrm{s}}M_*)^{2/3}, &\text{for}~\nu_1 M_*\le \nu_{\mathrm{s}}M_*<\nu_2M_*\\
                                                       (\nu_1M_*)^{-1}(\nu_2M_*)^{-4/3}\frac{(\sigma M_*)^{4}(\nu_{\mathrm{s}}M_*)^{2}}{\left(\left(\sigma M_*\right)^{2}+4\left(\nu_{\mathrm{s}} M_*-\nu_{2} M_*\right)^{2}\right)^{2}}, &\text{for}~\nu_2 M_*\le \nu_{\mathrm{s}}M_*\le \nu_3 M_*\\
                                                       0, &\text{for}~\nu_3 M_* < \nu_{\mathrm{s}} M_*
                                                   \end{cases}
    \end{aligned}
\end{equation}
where $\sigma M_*$ and $\nu_i M_*$ ($i=1,2,3$) are given by
\begin{align}
&\nu_1 M_*=(1-4.455+3.521)+0.6437\eta-0.05822\eta^2-7.092\eta^3 \frac{1}{\pi(s_1+s_2)},\\
&\nu_2 M_* = (1-0.63)/2+0.1469\eta-0.0249\eta^2+2.325\eta^3 \frac{1}{\pi(s_1+s_2)},\\
&\sigma M_* = (1-0.63)/4 -0.4098\eta +1.829\eta^2-2.87\eta^3 \frac{1}{\pi(s_1+s_2)},\\
&\nu_3 M_* = 0.3236 -0.1331\eta -0.2714\eta^2 +4.922\eta^3 \frac{1}{\pi(s_1+s_2)}.
\end{align}
Following Refs.~\cite{Raidal:2018bbj, Liu:2018ess, Hutsi:2020sol}, the merger rate of PBH binaries formed in the early universe is
\begin{equation} \label{R}
    R=\frac{1.6 \times 10^6}{\mathrm{Gpc}^3 \mathrm{yr}} f_{\mathrm{PBH}}^{\frac{53}{37}}\left(\frac{t}{t_0}\right)^{-\frac{34}{37}}\left(\frac{M_t}{M_{\odot}}\right)^{-\frac{32}{37}} \eta^{-\frac{34}{37}} S(\psi, f_{\mathrm{PBH}}, M_{t}) \psi\left(M_1\right) \psi\left(M_2\right)
\end{equation}
where$\psi(M) \equiv \frac{f(M)}{M f_{\mathrm{PBH}}}$ and $S(\psi, f_{\mathrm{PBH}}, M_t)$ is a suppression factor accounting for the effect of the surrounding smooth matter component (not PBHs) on the binary formation and the disruption of the binary by other PBHs.
Following Ref.~\cite{Liu:2018ess}, we adopt $S=\left(1+\sigma_{\mathrm{M}}^2 / f_{\mathrm{PBH}}^2\right)^{-21 / 74}$. The merger rate $RdM_1dM_2$ could be expressed as
\begin{equation}
    \begin{aligned}
        RdM_1dM_2&=\frac{1.6 \times 10^6}{\mathrm{Gpc}^3 \mathrm{yr}}\left(\frac{t}{t_0}\right)^{-\frac{34}{37}}\left(\frac{M_*}{M_{\odot}}\right)^{-\frac{32}{37}}(s_1+s_2)^{\frac{36}{37}}(s_1s_2)^{-\frac{71}{37}}f(s_1)f(s_2)f_{\mathrm{PBH}}^{-\frac{21}{37}} S ds_1ds_2 \\
                 &=\tilde{R}(z,s_{1},s_{2})ds_1ds_2
    \end{aligned}
\end{equation}
The spectrum of GWs produced from coalescing PBH binaries can be rewritten as
\begin{equation}
    \Omega_{\GW}^{\BB}(\nu M_*) = \frac{\nu M_*}{\rho_c M_*} \int \dif s_{1} \dif s_{2} \int_{0}^{\frac{\nu_{\rmc} M_*}{\nu M_*}-1} \dif z \frac{\tilde{R}(z,s_{1},s_{2})}{(1+z)H(z)} \tilde{E}(\nu_{\mathrm{s}}M_*,s_{1},s_{2}).
\end{equation}

\bibliographystyle{JHEP}
\bibliography{citeLib}

\end{document}